\documentclass[fleqn,10pt]{wlscirep}
\title{Improved estimation of the effective reproduction number with heterogeneous transmission rates and reporting delays}

\author[1,2]{Xin-Jian Xu}
\author[1]{Song-Jie He}
\author[3,*]{Li-Jie Zhang}
\affil[1]{Department of Mathematics, Shanghai University, Shanghai 200444, China}
\affil[2]{Qian Weichang College, Shanghai University, Shanghai 200444, China}
\affil[3]{Department of Physics, Shanghai University, Shanghai 200444, China}
\affil[*]{lijzhang@shu.edu.cn}

\begin{abstract}
In the face of an infectious disease, a key epidemiological measure is the basic reproduction number, which quantifies the average secondary infections caused by a single case in a susceptible population. In practice, the effective reproduction number, denoted as $R_t$, is widely used to assess the transmissibility of the disease at a given time $t$. Real-time estimating this metric is vital for understanding and managing disease outbreaks. Traditional statistical inference often relies on two assumptions. One is that samples are assumed to be drawn from a homogeneous population distribution, neglecting significant variations in individual transmission rates. The other is the ideal case reporting assumption, disregarding time delays between infection and reporting. In this paper, we thoroughly investigate these critical factors and assess their impact on estimating $R_t$. We first introduce negative binomial and Weibull distributions to characterize transmission rates and reporting delays, respectively, based on which observation and state equations are formulated. Then, we employ a Bayesian filtering for estimating $R_t$. Finally, validation using synthetic and empirical data demonstrates a significant improvement in estimation accuracy compared to conventional methods that ignore these factors.
\end{abstract}

\begin{document}

\flushbottom
\maketitle
%
%
\thispagestyle{empty}


\section*{Introduction}

When assessing an infectious disease outbreak, the available data of interest is the time series of new infection cases~\cite{cauchemez2006}. This dataset not only helps determine the impact and scale of the epidemic but also provides insights into trends and variations within its transmission dynamics~\cite{fraser2007}. These patterns are commonly depicted by effective reproduction numbers, symbolized as $R_t$ at time $t$, which quantify the average number of new infections caused by previously infectious individuals. In principle, an $R_t$ value greater than $1$ suggests a growing incidence rate, while a value less than $1$ indicates a declining epidemic~\cite{parag2020using}.
    
The reliable estimation of the time-varying trend of $R_t$ is a critical aspect of mathematical epidemiology. With the unfolding of the COVID-19 pandemic, there has been a growing interest in addressing this issue~\cite{thompson2020}. Studies have shown that $R_t$ plays a central role in enhancing individual awareness~\cite{bhatia2021} and providing valuable information for decision-making~\cite{anderson2020}. Initially, the focus was on understanding how changes in $R_t$ are linked with interventions such as lockdowns and social distancing~\cite{hao2020}. However, as countries transitioned to pandemic mitigation, the focus shifted to describing how to relax existing interventions with minimal risk~\cite{han2021}. In this context, estimating $R_t$ becomes increasingly significant as it serves as a key indicator describing the speed of virus transmission. Consequently, it aids decision-makers in better understanding the trends in epidemic propagation and adjusting prevention and control strategies promptly.

In the real world, however, individual variations are widespread, particularly regarding individual transmission rates and reporting delays, which have sparked extensive discussions~\cite{lloyd2005,althouse2020}. Transmission heterogeneity refers to variations in individual efficiency during the transmission process, and a comprehensive understanding of this property is crucial for implementing effective control measures. Given that a significant portion of transmission events is caused by a small number of individuals, targeting the identification and isolation of these \lq\lq super spreaders\rq\rq may prove more effective in disease containment~\cite{parag2021sub}. Reporting delays stem from differences in individuals' perception and reporting of events or diagnoses~\cite{li2021}. For an accurate estimation of $R_t$ at time $t$, it is imperative to consider and explicitly address delays from infection to reporting within a model.
    
Mathematically, there are various methods for estimating $R_t$. The classic approach is through differential equation models in structured population dynamics~\cite{barril2018,breda2021,ripoll2023}. For example, Kemp et al.~\cite{kemp2021} extended the standard SEIR model~\cite{kermack1927} to incorporate social interaction parameters, the presence of undetected cases, vaccination, and disease progression through hospitals, ICUs, recovery, and death. They used the next-generation matrix method to calculate $R_t$. By capturing population flows and infection rates, the trends can be predicted. However, the classic epidemic models possess certain limitations. Firstly, it relies on simplifying assumptions, like homogeneous mixing of the population and constant infection rates, which may not fully capture the complexity of interpersonal relationships and transmission dynamics in reality~\cite{guo2023}. Secondly, it overlooks individual differences, such as variations in immune levels, individual perceptions and behavioral patterns, potentially impacting the accurate characterization of transmission dynamics. 

Time series based methods offer a solution to these challenges. By their nature, time series models can effectively adapt to intricate data patterns and dynamic shifts, handling substantial amounts of observational data. This capability enhances the precision of disease transmission predictions. When estimating $R_t$ from a time series incidence rate curve, one can conceptualize the transmission process as a Poisson process. Under this assumption, the number of new infections $I_t$ at time $t$ can be computed by the following updating equation~\cite{fraser2007}:
\begin{equation}\label{updat}
I_t=\int_0^\infty\beta(t,\tau)I(t-\tau)d\tau,
\end{equation}
where $\beta(t,\tau)$ is the transmission rate from time $\tau$ to $t$. This updating model holds promise for real-time comprehension of the dynamics of an emerging epidemic. By assuming (i) homogeneous spread of the epidemic in the population, implying that the average number of infections per person at time $t$ is $R_{t}$, (ii) ideal case reporting, where all cases $I_{1}^{t}:=\{I_{1},I_{2},\ldots I_{t}\}$ are local cases in the monitored area, and (iii) $\Lambda_{t}=\sum_{u=1}^{t-1}I_{t-u}w_{u}$, where the distribution of intervals $w_{u}$ between sequences is known~\cite{wallinga2007}, one can obtain a Poisson distribution:
\begin{equation}\label{obeq}
I_t \sim \text{Poiss}(R_t\Lambda_t),
\end{equation}
which enables the real-time inference of $R_t$ from the observed sequence $I_{1}^{t}$ via the posterior distribution $P_t(R_{t}|I_{1}^{t})$. 

While the model assumes that the number of new infections on day $t$ can be depicted by a Poisson distribution, assumptions about the continuity of $R_t$ can broadly be categorized into two groups. The first category assumes that $R_t$ is piecewise constant within a predetermined sliding window. For example, Cori et al.~\cite{cori2013} devised a method to estimate $R_t$ from the updating equation, assuming $R_t$ remains constant within a sliding window of period $k$. However, the choice of $k$ impacts estimation, as large or small values of $k$ can lead to oversmoothing (ignoring significant changes) or misinterpreting random noise as meaningful. In view of this, Parag et al.~\cite{parag2020using} proposed a scheme to optimize the selection of $k$ using the accumulative prediction error based on information theory. Nonetheless, the fixed value of $k$ may fail to accurately identify change points in $R_t$, potentially treating distinct $R_t$ values as identical within the sliding window. More recently, Creswell et al.~\cite{creswell2023} introduced a Bayesian nonparametric model based on the Pitman-Yor process, which determines when or whether $R_t$ should change based on the rate curve and prior information, as well as how many changes should occur throughout the sequence of $I_{1}^{t}$. Despite providing valuable transmission estimates, these methods have limitations, particularly in case of the short sequence, where the estimation is heavily influenced by the prior distribution.

The second category posits that $R_t$ evolves smoothly with variations governed by a Gaussian filter. Given that $R_t$ fluctuates daily, it is expected to exhibit autocorrelation~\cite{parag2021deciphering}. Parag et al.~\cite{parag2021improved} introduced a novel approach, termed EpiFilter, which converts the updating equation~\eqref{updat} into a state-space model incorporating a Markov assumption. In this way, $R_t$ is regarded as a hidden state to be inferred. It dynamically relies on both the previous state $R_{t-1}$ and the noise term $\epsilon_{t-1}$:
\begin{equation}
  R_t=f_t(R_{t-1}, \epsilon_{t-1}),
\end{equation}
where the form of $f_t$ is to be determined. For simplicity, we assume that a noisy linear projection of states over consecutive timepoints provides a good approximation of the state trajectory~\cite{snyder1975}. We scale the noise of this projection by a fraction $\eta<1$ of the magnitude of $\sqrt{R_{t-1}}$. Consequently, a linear form for $f_t$ is obtained and the state equation can be rewritten as:
\begin{equation}\label{state}
 R_{t}=R_{t-1}+\left(\eta\sqrt{R_{t-1}}\right)\epsilon,
\end{equation}
where $\epsilon$ is  a standard normal distribution, $\epsilon \sim \mathrm{Norm}(0,1)$. 

The observation equation reads
\begin{equation}\label{obser}
 I_t=g_t(R_t, \sigma_t),
\end{equation}
where $\sigma_t$ is also a white noise. Eq.~\eqref{obser} describes the relationship between the observed value $I_t$ and the state value $R_t$ at time $t$. However, the explicit form of $g_t$ usually remains unknown, thus rendering it implicit. To address this issue, we assume that $g_t$ adheres to a specific probability distribution. In the case of homogeneous transmission, it conforms to Eq.~\eqref{obeq}. Then, one can compute $P(R_{t}|I_{1}^{t})$ through Bayesian filtering.

\begin{table}[ht]
  \centering
  \caption{Notation used in the paper.}\label{notion}
  \begin{tabular}{c|l}
    \toprule
    \textbf{Term} & \textbf{Meaning} \\ 
    \hline
    $R_t$ & Effective reproduction numbers  \\
    \hline
    $\beta(t,\tau)$ & The transmission rate from time $\tau$ to $t$ \\
    \hline
    $I_t$ & The report number of new infections at time $t$ \\
    \hline
    $A_t$ & The actual number of infections at time $t$ \\
    \hline
    $\Lambda_{t}$ & Total infectivity \\
    \hline
    $w_{u}$ & Serial interval distribution \\ 
    \hline
    $k$ & Sliding window of period  \\
    \hline
    $\epsilon_{t}$,$\epsilon^{\prime}$,$\epsilon^{\prime\prime}$ & White noise  \\
    \hline
    $d_u$ & Delay distribution \\
    \hline
    $\eta$,$\eta_R$ & Correlation parameters of the effective reproduction number \\
    \hline
    $\eta_r$ & Correlation parameter of dispersion \\
    \hline
    $\mathcal{R}$ & Closed space representing valid values of $R$ \\
    \hline
    $\delta_R$ & Grid size of $R$  \\
    \hline
    $m_R$ & Partition for $R$  \\
    \hline
    $R_{\mathrm{min}}$ & Minimum value of $R$  \\
    \hline
    $R_{\mathrm{max}}$ & Maximum value of $R$  \\
    \hline
    $\Upsilon$ & Closed space representing valid values of $r$  \\
    \hline
    $\delta_r$ & Grid size of $r$  \\
    \hline
    $m_r$ & Partition for $r$  \\
    \hline
    $r_{\mathrm{min}}$ & Minimum value of $r$  \\
    \hline
    $r_{\mathrm{max}}$ & Maximum value of $r$  \\
    \bottomrule
  \end{tabular}
\end{table}

Both of these methods rely on the updating equation, which is built upon two key assumptions. Firstly, they assume homogeneity in the spread of an epidemic throughout a population, implying that the number of secondary cases generated by each individual conforms well to a Poisson distribution. However, epidemic spread within the population often varies in terms of individual transmission rates in the real world~\cite{gostic2020,kuylen2022}, indicating differences in the dispersion of the number of offspring. Since the Poisson distribution has identical mean and variance, it often fails to adequately describe the diverse dispersion levels of the offspring count. In this paper, we choose the negative binomial offspring distribution to model the daily count of new infection cases $I_t$, as it offers a better depiction of the secondary cases caused by each individual. Secondly, the assumption of ideal case reporting is made. However, in reality, the reported cases often stem from several days later than solely from the current day, suggesting the presence of a reporting delay. This delay poses challenges in obtaining comprehensive reports. Notably, in an epidemiological distribution study, Hawryluk et al.~\cite{hawryluk2020} proposed a joint Bayesian method for fitting and found that the delay distribution can be well fitted by the Weibull distribution. In contrast, Bizzotto et al.~\cite{bizzotto2023} introduced a consolidation function that relies on the number of cases reported for a specific symptom onset date across successive reporting updates, in order to characterize the reporting delay on that particular day. In this study, we adopt the Weibull distribution to characterize the varying time from case occurrence to reporting.

By incorporating these two distributions, we enhance the EpiFilter to provide a more precise representation of the complex dynamics of epidemic transmission including individual variations and reporting delays, called the generalized EpiFilter. Through synthetic data, we demonstrate the effectiveness of our approach in accurately tracking changes in $R_t$. Using the Mean Squared Error (MSE), we assess the accuracy of our estimations, revealing that the generalized EpiFilter yields smaller MSE values compared to the standard EpiFilter. Furthermore, we apply the proposed framework to the COVID-19 incidence curves in New York State and New Zealand to showcase its practical utility. Our results indicate a significant improvement in $R_t$ estimate, closely aligning with locally implemented intervention strategies.

\section*{Methods}

The observation equation establishes the connection between the hidden variable $R_t$ and the observable data $I_t$. Here, we explore observation equations for three distinct scenarios: (i) accounting for individual heterogeneity in transmission rates while disregarding reporting delays; (ii) incorporating individual reporting delays while neglecting transmission variability; and (iii) simultaneously addressing both heterogeneous rates and reporting delays. All notations are summarised in Table~\ref{notion} for reader convenience.

\subsection*{The observation equation incorporating transmission heterogeneity without reporting delays}

In models concerning offspring distribution, a Poisson distribution is commonly used to approximate the variable of interest. While the Poisson distribution assumes identical mean and variance, making it unsuitable for data reflecting individual heterogeneity in transmission rates. As an alternative, the negative binomial distribution $\text{NB}(n,p)$ consistently exhibits a variance exceeding the mean. Here, $1/n$ signifies the dispersion parameter. This characteristic endows the negative binomial distribution with greater versatility, making it better suited for a wider range of practical applications by accommodating dispersion, particularly individual heterogeneity in transmission rates~\cite{williams2024}. Consequently, we posit that the observed number $I_t$ of reported cases is generated by a negative binomial model, i.e.,
\begin{equation}\label{nbit}
I_t\sim \text{NB}(\mu_t,r_t),
\end{equation}
where $\mu_t$ and $r_t$ represent mean and dispersion at time $t$, respectively. The corresponding probability density function is given by \begin{equation}
p(I_t=k)=\frac{\Gamma(r_t+k)}{k!\Gamma(r_t)}\left(\frac{r_t}{r_t+\mu_t}\right)^{r_t}\left(\frac{\mu_t}{r_t+\mu_t}\right)^k.
\end{equation}
Since the actual number $A_t$ of infections each day is defined as
\begin{equation}\label{defat}
A_t=R_t\Lambda_{t},
\end{equation}
we have $\mu_t=A_t=R_t\Lambda_{t}$. Substituting it into Eq.~\eqref{nbit} yields the observation equation:
\begin{equation}\label{ob1}
I_t\mid R_{\tau},r_{t}\sim \text{NB}(R_t\Lambda_{t},r_t),
\end{equation}
where ($\cdot \mid \cdot$) represent the conditional distribution.

\subsection*{The observation equation incorporating reporting delays without transmission heterogeneity}

Assuming a delay distribution denoted with $d_u$, where $d_u$ signifies the probability that cases infected on day $t-u+1$ are reported on day $t$, the observed $I_t$ is derived from infections $A_{\tau}$ occurring in the preceding $\tau$ days under the delay distribution $d_u$. Consequently, the expected value of $I_t$ can be written as
\begin{equation}\label{expit}
\mu_t=\sum_{\tau}d_{t-\tau}A_{\tau}.
\end{equation}
Substituting Eq.~\eqref{defat} into Eq.~\eqref{expit} yields $\mu_t=\sum_{\tau}d_{t-\tau}R_{\tau}\Lambda_{\tau}$. Because of no heterogeneity in individual transmission, the number of new infections follows a Poisson distribution:
\begin{equation}\label{itpoiss1}
I_t\sim \text{Poiss}(\mu_t).
\end{equation}
Substituting Eq.~\eqref{expit} into Eq.~\eqref{itpoiss1} yields
\begin{equation}\label{itdelay}
I_t\mid R_{\tau}\sim \text{Poiss}\left(\sum_{\tau}d_{t-\tau}R_{\tau}\Lambda_{\tau}\right).
\end{equation}
Assuming the current $I_t$ solely depending on the current $R_{t}$, the above equation can be rewritten as
\begin{equation}
I_t\mid R_{\tau} \sim \text{Poiss}\left(d_{0}R_{t}\Lambda_{t}+\sum_{\tau^{\prime}}d_{t-\tau^{\prime}}R_{\tau^{\prime}}\Lambda_{\tau^{\prime}}\right),
\end{equation}
where $\tau^{\prime}$ represents $\tau$ excluding $t$. Replacing $R_{\tau^{\prime}}$ by the estimate $\hat{R}_{\tau^{\prime}}$, we obtain the observation equation:
\begin{equation}\label{ob2}
I_t\mid R_{t}\sim \text{Poiss}\left(d_{0}R_{t}\Lambda_{t}+\Delta\right),
\end{equation}
where $\Delta=\sum_{\tau^{\prime}}d_{t-\tau^{\prime}}\hat{R}_{\tau^{\prime}}\Lambda_{\tau^{\prime}}$.

\subsection*{The observation equation incorporating both factors}

Substituting Eq.~\eqref{expit} into Eq.~\eqref{nbit} yields
\begin{equation}
I_t\mid R_{\tau},r_{t}\sim \text{NB}\left(\sum_{\tau}d_{t-\tau}R_{\tau}\Lambda_{\tau},r_t\right).
\end{equation}
Using the same assumption that the current $I_t$ depends only on the current $R_{t}$, we rewrite the above equation as
\begin{equation}
I_t\mid R_{\tau},r_{t}\sim \text{NB}\left(d_{0}R_{t}\Lambda_{t}+\sum_{\tau^{\prime}}d_{t-\tau^{\prime}}R_{\tau^{\prime}}\Lambda_{\tau^{\prime}},r_t\right),
\end{equation}
where $\tau^{\prime}$ represents $\tau$ excluding $t$. Finally, replacing $R_{\tau^{\prime}}$ by the estimate $\hat{R}_{\tau^{\prime}}$, we obtain the observation equation:
\begin{equation}\label{ob3}
I_t\mid R_{t},r_{t}\sim NB(d_{0}R_{t}\Lambda_{t}+\Delta,r_t)
\end{equation}
where $\Delta=\sum_{\tau^{\prime}}d_{t-\tau^{\prime}}\hat{R}_{\tau^{\prime}}\Lambda_{\tau^{\prime}}$.

\subsection*{The state equation incorporating transmission heterogeneity}

Eq.~\eqref{state} characterizes the state equation without transmission heterogeneity. When accounting for this factor, it becomes essential to incorporate the offspring dispersion $r_t$, which quantifies the variability in the number of secondary infections generated by a single infected individual over a specific period. In our methodology, we adopt three modest assumptions. Firstly, we assert that the variability in offspring dispersion $r_t$ is only weakly influenced by $R_t$~\cite{susswein2020}, treating $R_{t}$ and $r_t$ as two independent states. Secondly, we presume that the noisy linear projection of the state at continuous time points offers a good approximation of the state trajectory~\cite{parag2021improved}. Thirdly, we introduce two parameters, $\eta_R$ and $\eta_r$, which govern the autocorrelations of the effective reproduction number $R_{t}$ and offspring dispersion $r_t$, respectively. Consequently, the state equation can be written as
\begin{equation}\label{stateq}
\begin{pmatrix}R_{t}\\r_{t}\end{pmatrix}=
\begin{pmatrix}R_{t-1}\\r_{t-1}\end{pmatrix}
+\begin{pmatrix}\eta_R\sqrt{R_{t-1}}\epsilon^{\prime}\\
\eta_r\sqrt{r_{t-1}}\epsilon^{\prime\prime}\end{pmatrix}.
\end{equation}
Typically, it is assumed that the two noises $\epsilon^{\prime}$ and $\epsilon^{\prime\prime}$ are independent of each other. In this way, we assume that they follow a two-dimensional standard normal distribution, $(\epsilon^{\prime},\epsilon^{\prime\prime}) \sim \mathrm{Norm}(0,0,0,1,1)$. 

\subsection*{Particle filtering for the state-space model}
 
In the state-space model, the process of estimating the hidden state at time $t$ using the preceding $t$ observations is termed filtering, expressed as $P_{t}(R_{t},r_{t}|I_{1}^{t})$. Typically, this is tackled through Bayesian filtering techniques, such as Kalman filtering and particle filtering. Given the non-Gaussian nature of our observation equation, we employ particle filtering to compute $P_{t}(R_{t},r_{t}|I_{1}^{t})$, which operates by utilizing a set of random samples (particles) along with their associated weights to depict the posterior distribution of the state. This method involves selecting an probability density function and performing random sampling from it~\cite{chen2003}. After obtaining these random samples and their corresponding weights, adjustments are made based on the state observations to refine the weights and positions of the particles. These samples are then utilized to approximate the posterior distribution of the state, with the weighted sum serving as the state estimate. Unlike methods constrained by linear and Gaussian assumptions, particle filtering describes the state probability density via sample representation, rather than functional representation, thus avoiding undue constraints on the probability distribution of the state variables.

For $R_{t}$, we establish a closed space $\mathcal{R}$ representing valid values of $R$~\cite{parag2021improved}. Given a partition $m_R$, extreme values $R_{\mathrm{min}}$ and $R_{\mathrm{max}}$, and grid size $\delta_R=(R_{\mathrm{max}}-R_{\mathrm{min}})/m_R$, we define $\mathcal{R}$ as $\{R_{\mathrm{min}},R_{\mathrm{min}}+\delta_R,\ldots,R_{\mathrm{max}}\}$. This indicates that the instantaneous reproduction number $R_{t}$ must take a discrete value within $\mathcal{R}$, with its $i$th element denoted as ${\mathcal{R}}[i]$. Similarly, for $r_{t}$, we establish a closed space $\Upsilon$ representing valid values of $r$. For a given partition $m_r$, extreme values $r_{\mathrm{min}}$ and $r_{\mathrm{max}}$, and grid size $\delta_r=(r_{\mathrm{max}}-r_{\mathrm{min}})/m_r$, we define $\Upsilon$ as $\{r_{\mathrm{min}},r_{\mathrm{min}}+\delta_r,\ldots,r_{\mathrm{max}}\}$. This means that the parameter $r_{t}$ must take a discrete value within $\Upsilon$, with its $i$th element denoted as ${\Upsilon}[i]$. We formalize these two variables in the following equations:
\begin{eqnarray}
\sum_{i=1}^{m}p(R_{s}=\mathcal{R}[i])&=&1, \quad 1\leq s\leq t,\\
\sum_{i=1}^{m}p(r_{s}=\Upsilon[i])&=&1, \quad 1\leq s\leq t.
\end{eqnarray}

Particle filtering comprises two primary steps: prediction and update. Initially, the prediction step computes the prior predictive distribution:
\begin{equation}\label{filter1}
p_t=\mathbb{P}(R_{t},r_{t}|I_{1}^{t-1})
=\iint \mathbb{P}(R_{t},r_{t}\mid R_{t-1},r_{t-1},I_{1}^{t-1})P_{t-1}\text{d}R_{t-1}\text{d}r_{t-1},
\end{equation}
leveraging historical incidence rate data $I_{1}^{t-1}$ and the previous state $R_{t-1}$. This is also known as propagation. Here, $\mathbb{P}(R_{t},r_{t}\mid R_{t-1},r_{t-1},I_{1}^{t-1}) \sim N(R_{t-1},r_{t-1},\eta_R^{2}R_{t-1},\eta_r^{2}r_{t-1},0)$ is the state model from Eq.~\eqref{state} or Eq.~\eqref{stateq}. In the update step, this prior prediction undergoes adjustment to yield the posterior filtering distribution:
\begin{equation}\label{filter2}
P_{t}=\mathbb{P}(R_{t},r_{t}|I_{1}^{t})
=\frac{\mathbb{P}(R_{t},r_{t},I_{t}|I_{1}^{t-1})}{\mathbb{P}(I_{t}|I_{1}^{t-1})}
=C^{-1}\mathbb{P}(I_{t}|R_{t},r_{t},I_{1}^{t-1})p_t,
\end{equation}
where $C=\iint \mathbb{P}(I_{t}|R_{t},r_{t})\mathbb{P}(R_{t},r_{t}|I_{1}^{t-1})\text{d}R_{t}\text{d}r_{t}$ is the normalization and $\mathbb{P}(I_{t}|R_{t},r_{t},I_{1}^{t-1})$ is the observation model from Eq.~\eqref{ob1}, Eq.~\eqref{ob2} or Eq.~\eqref{ob3}. This is also known as correction. Solving Eq.~\eqref{filter2} yields
\begin{equation}\label{rtdist}
\mathbb{P}(R_{t}|I_{1}^{t})=\int\mathbb{P}(R_{t},r_{t}|I_{1}^{t})\text{d}r_t.
\end{equation}
However, obtaining the exact integrals in Eq.~\eqref{filter1} and Eq.~\eqref{filter2} is challenging. To circumvent this difficulty, we utilizes a collection of random samples, along with their associated weights to approximate the posterior distribution of the system's state. We commence this procedure by initializing a two-dimensional uniform prior distribution over the space of $(\mathcal{R},\Upsilon)$ for $P_1$. It is important to note that $p_t$ and $P_t$ are $m_R*m_r$ matrices whose elements sum to $1$, with the ($i,j$)-th entry representing the scenario where $(R_t,r_t)=(\mathcal{R}[i],\Upsilon[j])$. By iterating over the grid of ($\mathcal{R},\Upsilon$) while concurrently solving Eq.~\eqref{filter1} and Eq.~\eqref{filter2}, we obtain the real-time estimate of $R_{t}$.

\section*{Results}

\subsection*{Estimating $R_t$ with transmission heterogeneity}

During the global spread of COVID-19, we once again witness instances of super-spreading events, wherein specific individuals infect a substantial number of secondary cases~\cite{zhang2020}. Given that variations in the effective reproduction number reflect changes in transmission and incidence rates, accurately identifying and interpreting trends in $R_t$ within heterogeneous transmission scenarios is paramount for effective prevention and timely epidemic interventions. In this section, we compare observation equations that account for heterogeneity with those that don't.

\begin{figure*}[ht]
 \centering
 \includegraphics[scale=0.25]{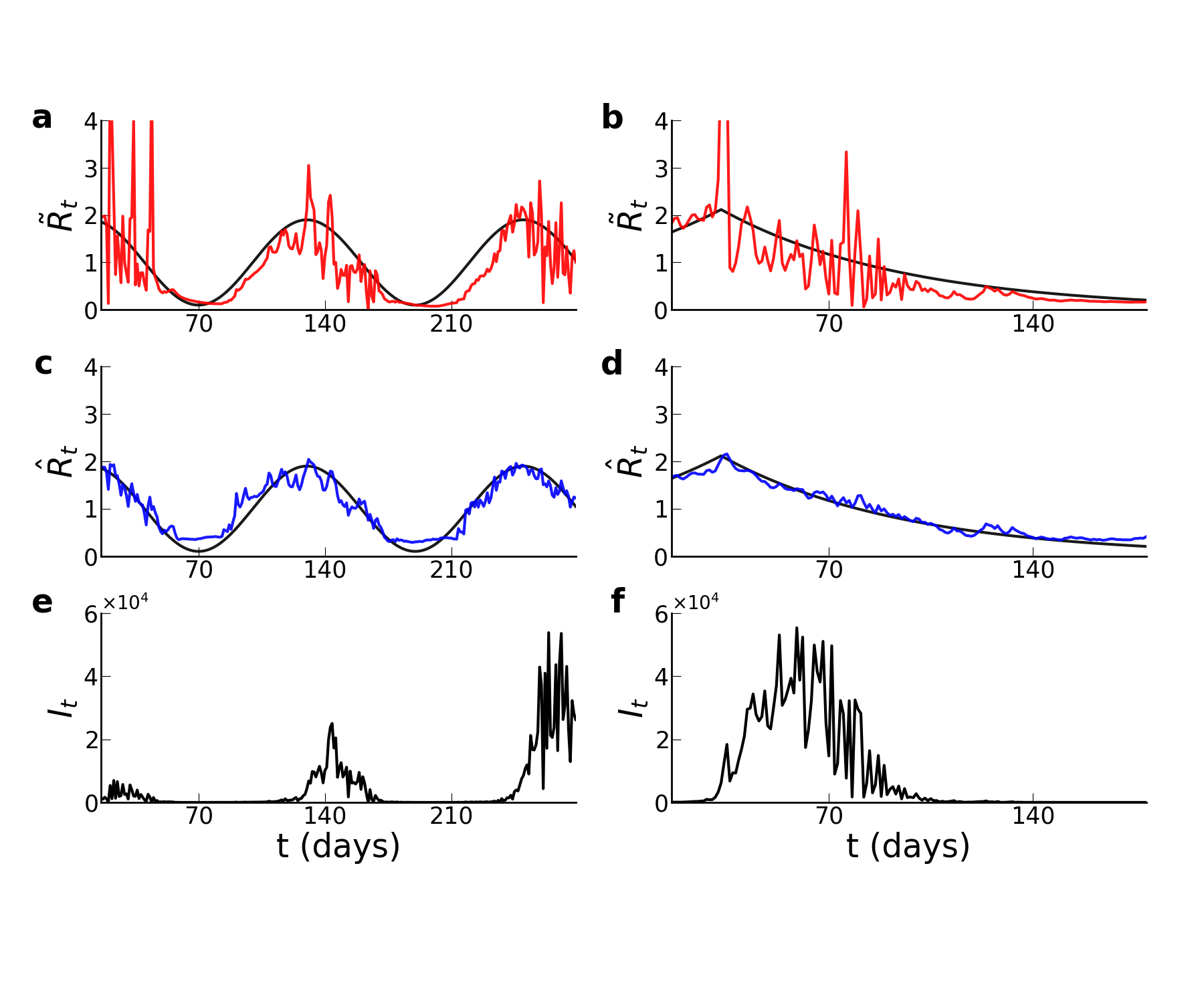}
 \vspace{-1.5cm}
 \caption{Estimates of the effective reproduction number with individual transmission heterogeneity in recurring outbreaks (left panel) and an initial outbreak followed by a gradual decline towards elimination (right panel). The true values of $R_t$ and $I_t$ are depicted in black. The red and blue lines represent $\tilde{R}_t$ and $\hat{R}_t$, corresponding to the estimates of the effective reproduction number obtained by the standard and generalized Epifliters, respectively. These estimates are implemented using a grid with $m=1000$, $R_{\text{min}}=0.01$ and $R_{\text{max}}=5$.}
 \label{heterest} 
\end{figure*}

In the context of transmission heterogeneity, the data generation process is as follows: (i) generate $R_t$ and $r_t$ curves for various scenarios using their respective initial values of $R_1$ and $r_1$ (black lines in Figs.~\ref{heterest}(a-d)); (ii) adopt the serial interval distribution $w_u$, which follows a gamma distribution with a mean of 4.8 and a variance of 5.29~\cite{nishiura2020}; (iii) set the initial values $\Lambda_1=0$ and $I_1=10$; (iv) calculate $\Lambda_2$ based on $\Lambda_{t}=\sum_{u=1}^{t-1}I_{t-u}w_{u}$; and (v) input $R_2$ and $r_2$ into Eq.~\eqref{nbit} to calculate $I_2$. Repeat steps (iv) and (v) to compile a comprehensive dataset for $I_t$ under transmission heterogeneity (black lines in Fig.~\ref{heterest}(e) and Fig.~\ref{heterest}(f)). We investigate two distinct situations: (i) recurring outbreaks, assumed to exhibit periodic or seasonal transmission (modeled by $R_t$ as a sine curve with an amplitude of $0.8$ and a period of $120$ time units); and (ii) an initial outbreak followed by a gradual decline towards elimination (initial exponential rise, transitioning into a slow decline at $t=50$). For each case, we simulate $100$ infection curves and employ both the standard EpiFilter (with $\eta=0.1$) and our generalization to estimate $R_t$. Here, the parameter $\eta$ controls the correlation between $R_{t-1}$ and $R_t$. An $\eta$ that is too small may result in underfitting of the model, whereas an $\eta$ that is too large may result in overfitting. We set $\eta=0.1$ as it performs well in multiple epidemiological scenarios and is effective in automatically detecting transmission changes~\cite{parag2021improved}.

In the standard EpiFilter, we compute the estimated values $\tilde{R}_t$ of the effective reproduction number using Eq.~\eqref{obeq} and Eq.~\eqref{state}. Conversely, for our generalized framework that accounts for individual transmission heterogeneity, we compute the estimated values $\hat{R}_t$ of the effective reproduction number based on Eq.~\eqref{ob1} and Eq.~\eqref{stateq}. Figure~\ref{heterest} illustrates the results from one realization. When confronted with data reflecting such heterogeneity (see Fig.~\ref{heterest}(e) and Fig.~\ref{heterest}(f)), the accuracy of estimates by the standard EpiFilter is compromised in both situations, manifested by apparent overestimation or underestimation of $R_t$ (red lines in Fig.~\ref{heterest}(a) and Fig.~\ref{heterest}(b)). In contrast, our generalized EpiFilter effectively addresses this issue. As depicted in Fig.~\ref{heterest}(c) and Fig.~\ref{heterest}(d), our estimation (blue lines) maintains stability and precision. In our approach, the number of new infections, $I_t$, is not solely determined by $R_t$; instead, it is influenced by both $R_t$ and $\Lambda_t$. When accounting for data heterogeneity, we employ the negative binomial distribution to model $I_t$. Consequently, $I_1^{t-1}$ may exhibit considerable variation, which subsequently impacts $\Lambda_t$. Therefore, when $R_t$ is close to 1, the volatility may cause visible fluctuations. In contrast, substantial deviations of $R_t$ from 1 result in behavior that conforms to standard epidemiological expectations.

\begin{figure*}[ht]
 \centering
 \includegraphics[scale=0.25]{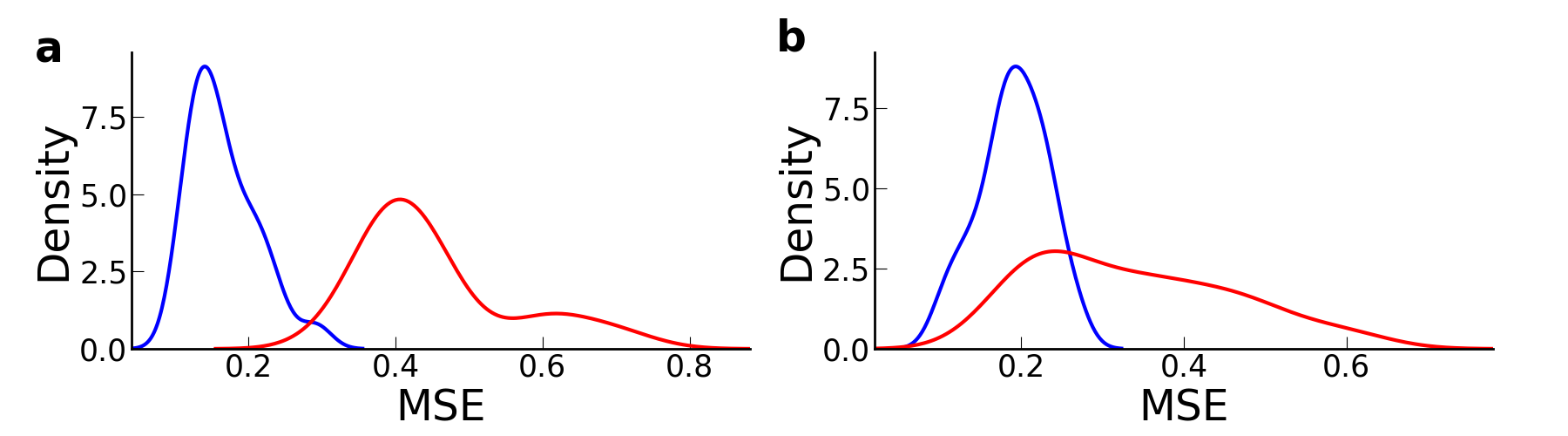}
 \caption{MSE distributions of the estimates derived from 100 realizations in recurring outbreaks (a) and an initial outbreak followed by a gradual decline towards elimination (b). The red and blue lines correspond to the standard and generalized EpiFilters, respectively. It is evident that our approach yields significantly smaller MSE values, with reductions of 3 to 4 times.}
 \label{hetermse}  
\end{figure*}

Additionally, we present the distributions of the MSE of $\tilde{R}_t$ and $\hat{R}_t$ over $100$ runs corresponding to two distinct situations in Fig.~\ref{hetermse}. Given a set of actual values $x_{i}$ and their estimated values $\hat{x}_{i}$, $i=1,2,\cdots,n$, the MSE is 
defined as
\begin{equation}
\mathrm{MSE}=\frac{1}{n}\sum_{i=1}^{n}\left(x_{i}-\hat{x}_{i}\right)^{2},
\end{equation}
which evaluates the proximity of estimated values to actual values. Substituting all $R_{i}$ and their estimated values $\tilde{R}_i$ or $\hat{R}_{i}$ into the above equation yields the MSE of the standard EpiFilter or its generalization. The lower the MSE, the closer the model's estimation is to the true value. To mitigate the randomness inherent in the sequence of $I_{t}$, we perform $100$ iterations of the simulation for both methods and obtain the MSE distributions. In the situation of recurring outbreaks (see Fig.~\ref{hetermse}(a)), the generalized EpiFilter exhibits the MSE distribution of with mean $0.01$ and variance $0.002$, contrasting with the standard EpiFilter's MSE distribution with mean $0.4$ and variance $0.02$. In the situation of an initial outbreak followed by a gradual decline towards elimination (see Fig.~\ref{hetermse}(b)), the generalized EpiFilter showcases the MSE distribution with mean $0.1$ and variance $0.01$, while the standard EpiFilter presents the MSE distribution with mean $0.5$ and variance $0.09$. These results underscore the superior accuracy and robustness of our improved framework in estimating $R_t$.

\begin{figure*}[ht]
 \centering
 \includegraphics[scale=0.25]{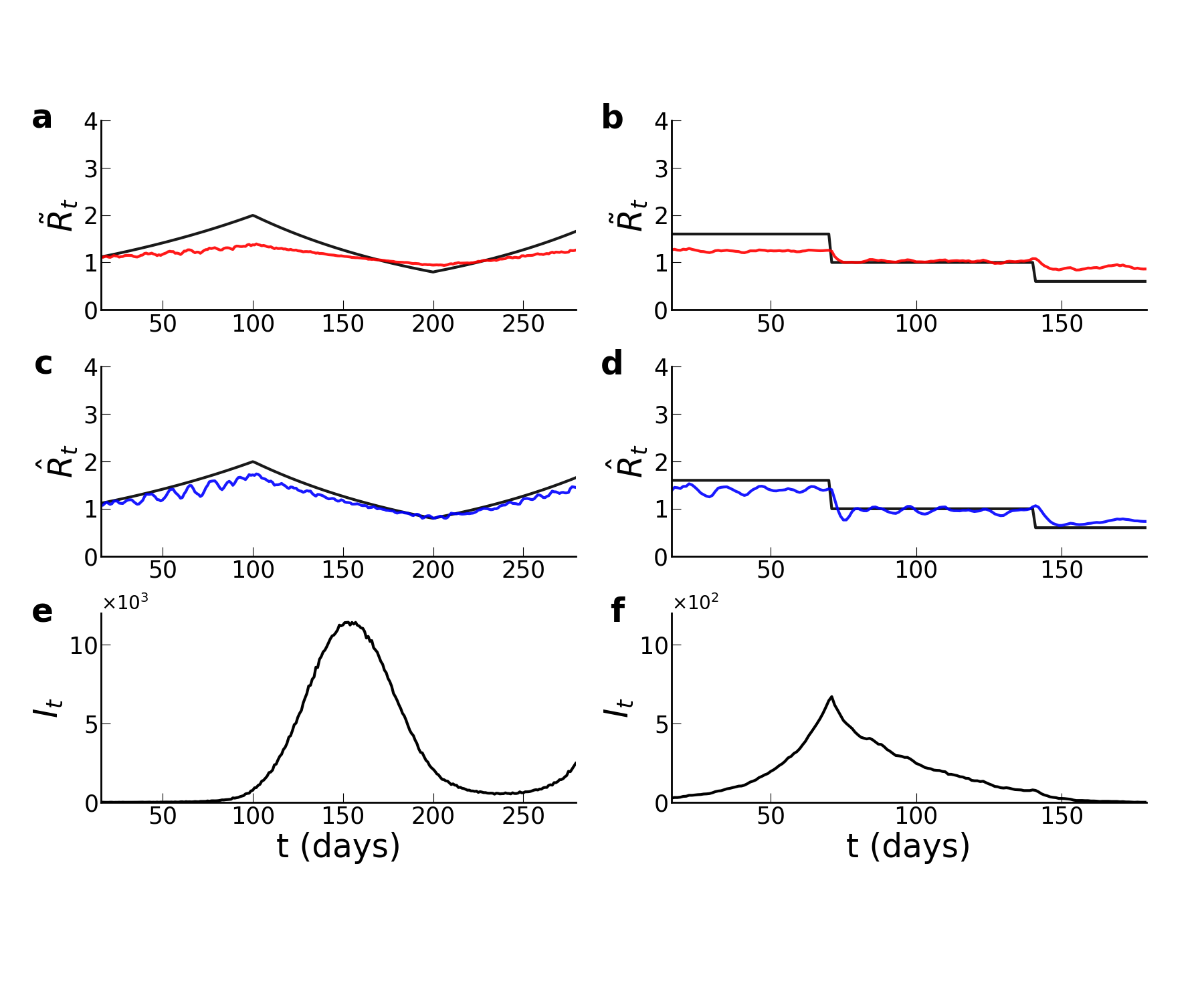}
 \vspace{-1.5cm}
 \caption{Estimates of the effective reproduction number with individual reporting delays in a rapid resurgence (left panel) and a gradual controlled epidemic (right panel). The true $R_t$ and $I_t$ are in black. The red and blue lines represent $\tilde{R}_t$ and $\hat{R}_t$, corresponding to the estimates of the effective reproduction number obtained by the standard and generalized Epifliters, respectively, using a grid with $m=1000$, $R_{\text{min}}=0.01$ and $R_{\text{max}}=5$.}
 \label{delayest}
\end{figure*}

\begin{figure*}[ht]
 \centering
 \includegraphics[scale=0.25]{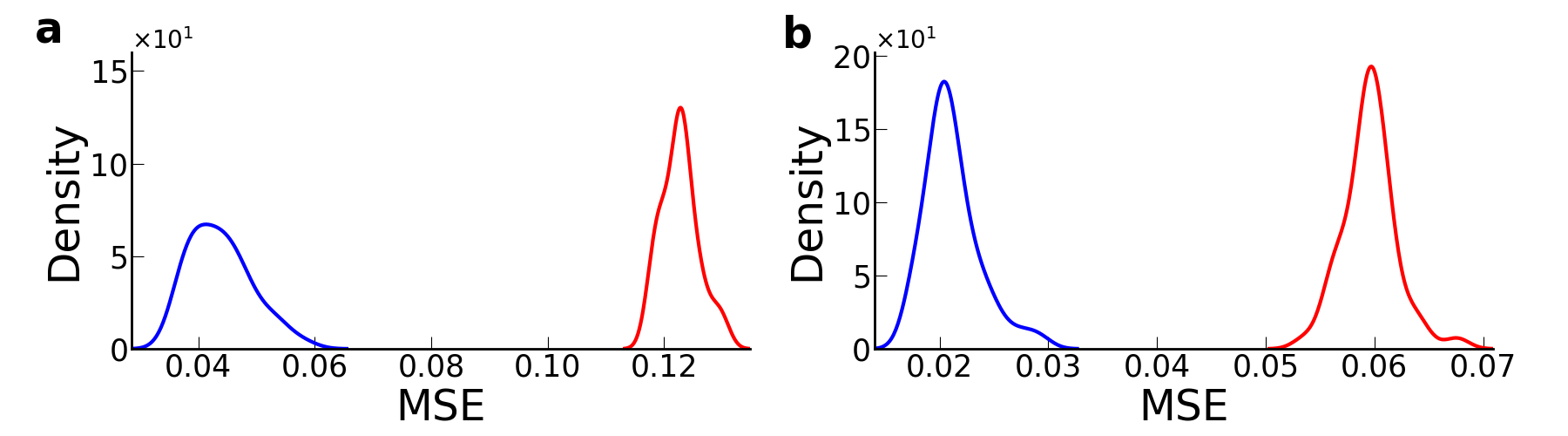}
 \caption{MSE distributions of the estimates from 100 realizations in a rapid resurgence (a) and a gradual controlled epidemic (b). The red and blue lines corresponding to the standard and generalized EpiFilters, respectively.  It is clear that our method results in significantly lower MSE values, achieving reductions of 2 to 3 times.}
 \label{delaymse}  
\end{figure*}

\subsection*{Estimating $R_t$ with reporting delays}

In the context of reporting delays, accurately estimating $R_t$ becomes pivotal for comprehending disease transmission dynamics. Especially during prolonged epidemics, it is essential to account for delays in recent case reporting to predict the onset of new waves effectively~\cite{harris2022}. In this section, we assess observation equations both with and without adjustments for these reporting delays in data-delayed scenarios. The data generation process is detailed as follows: (i) generate $R_t$ curves for various scenarios with their respective initial values of $R_1$ (black lines in Fig.~\ref{delayest}(a-d)); (ii) employ the serial interval distribution $w_u$, which follows a gamma distribution with a mean of 4.8 and a variance of 5.29~\cite{nishiura2020}, and utilize the Weibull distribution~\cite{hawryluk2020} to model the delay distribution $d_u$; (iii) set the initial values as $\Lambda_1=0$, $I_1=10$, and $A_1=I_1d_1$; (iv) calculate $\Lambda_2$ based on $\Lambda_{t}=\sum_{u=1}^{t-1}I_{t-u}w_{u}$; (v) submit $R_2$ and $\Lambda_2$ into Eq.~\eqref{obeq} to derive the actual values of $I_2$ without delay, denoted as $A_2$; (vi) input $A_1$ and $A_2$ into Eq.~\eqref{expit} to obtain $I_2$ with delay. Repeat steps (iv) and (vi) to construct a complete dataset for $I_t$ under reporting delays (black lines in Fig.~\ref{delayest}(e) and Fig.~\ref{delayest}(f)).  

We study two different situations: (i) an initially controlled epidemic, characterized by $R_t$ decreasing from $2$ to $0.8$ at $t=70$ succeeded by a rapid resurgence where $R_t$ rebounds from $0.8$ to $2$; and (ii) a gradual controlled epidemic, wherein $R_t$ decreases from $1.6$ to $1$ at $t=100$, followed by a further decline from $1$ to $0.6$ at $t=200$. For each case, we simulate $100$ infection curves and employ both the standard EpiFilter (with $\eta=0.1$) and our improved approach to estimate $R_t$. In the standard EpiFilter, we compute the estimate $\tilde{R}_t$ of the effective reproduction number using Eq.~\eqref{obeq} and Eq.~\eqref{state}. Conversely, for our generalization, we derive the estimate $\hat{R}_t$ of the effective reproduction number based on Eq.~\eqref{ob2} and Eq.~\eqref{stateq}.

Results from one realization are illustrated in Fig.~\ref{delayest}. Given the datasets incorporating reporting delays (black lines in Fig.~\ref{delayest}(e) and Fig.~\ref{delayest}(f)), the accuracy of the standard EpiFilter's estimates notably declines, as demonstrated in Fig.~\ref{delayest}(a) and Fig.~\ref{delayest}(b), revealing significant instances of overestimation or underestimation. When an epidemic outbreaks, delayed reporting significantly impacts the reported number of cases $I_t$ for a specific day, derived from infection data $A_{\tau}$ spanning the previous $\tau$ days, as governed by the delay distribution $d_u$. As a result, $I_t$ may fall below the true value $A_t$ for that day, potentially leading to an underestimation of $R_t$. While the epidemic diminishes, delayed reporting may cause $I_t$ to exceed the true value $A_t$ for a given day, potentially leading to an overestimation of the effective reproduction number. In contrast, our findings demonstrate that incorporating reporting delays results in a more accurate estimate of $R_t$, thus reducing the degrees of underestimation and overestimation. Moreover, Figure~\ref{delayest}(d) emphasizes that our generalized EpiFilter displays sharper transitions when $R_t$ experiences sudden changes, whereas the standard EpiFilter exhibits smoother transitions in estimating $R_t$ (see Fig.~\ref{delayest}(b)).

Furthermore, we present the distribution of the MSE for $\tilde{R}_t$ and $\hat{R}_t$ across $100$ runs in two distinct situations. In situation (i), our method yields the MSE of mean $0.04$ and variance $0.0002$, while the EpiFilter method results in the MSE of mean $0.124$ and variance $0.0009$ (see Fig.~\ref{delaymse}(a)). In situation (ii), our method induces the MSE of mean $0.02$ and variance $0.006$, whereas the EpiFilter method exhibits the MSE of mean $0.05$ and variance $0.006$ (see Fig.~\ref{delaymse}(b)). These findings highlight how our method strongly minimizes estimation errors in $R_t$ estimates produced by our method.

\subsection*{Estimating $R_t$ from empirical data}

In real-world scenarios, alongside the influence of transmission heterogeneity, reporting delays are also common. The delay distribution $D$ describes the duration from infection to reporting for an individual, a parameter difficult to accurately ascertain in practical settings. However, the reporting time closely correlates with the diagnosis time, allowing us to approximate our delay distribution with the duration from infection to diagnosis distribution. Consequently, our observation equation adopts Eq.~\eqref{ob3}.

We first analyze data collected from March 3rd, 2020 (the initial reported date) to September 15th, 2020, sourced from New York State, where reporting delays were notably observed, particularly in New York City~\cite{harris2022}. The serial interval distribution was best fitted using the gamma distribution with mean $4.8$ and variance $5.29$~\cite{nishiura2020}. Moreover, the distribution from infection to diagnosis was covered by the Weibull distribution with mean $4.8$ and variance $9.18$~\cite{kraemer2020}.

\begin{figure*}[ht]
  \centering
 \includegraphics[scale=0.25]{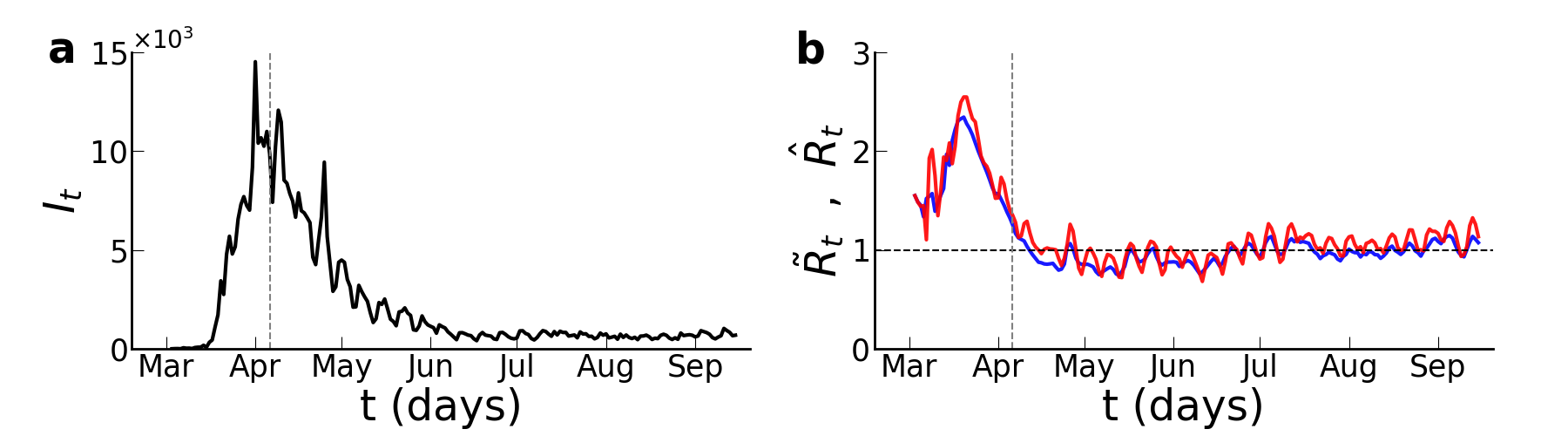}
  \caption{Temporal evolution of new infections of the COVID-19 in New York (a) and their estimates by the EpiFilter method (red line) and our approach (blue line).}
  \label{realny}
\end{figure*}

We compare the estimates of the effective reproduction number obtained from the EpiFilter method ($\tilde{R}_t$) and our approach ($\hat{R}_t$) in Fig.~\ref{realny}. These estimates were computed using Eq.~\eqref{obeq} and Eq.~\eqref{state} for the EpiFilter method, and Eq.~\eqref{ob3} and Eq.~\eqref{stateq} for our approach. Significantly, we observe notable differences in the inference quality between $\tilde{R}_t$ and $\hat{R}_t$. The estimates from the generalized Epifilter exhibit smaller fluctuations compared to the standard Epifilter, resulting in high accuracy. For instance, examining the period from July 1st to September 1st, we observe relatively stable daily new case counts (refer to Fig.~\ref{realny}(a)) with minor fluctuations. Consequently, the effective reproduction number during this interval is expected to remain relatively stable, hovering around $1$. Our estimates, accounting for transmission heterogeneity and reporting delays, closely approximate the true value (refer to Fig.~\ref{realny}(b)). 

\begin{figure*}[ht]
  \centering
 \includegraphics[scale=0.25]{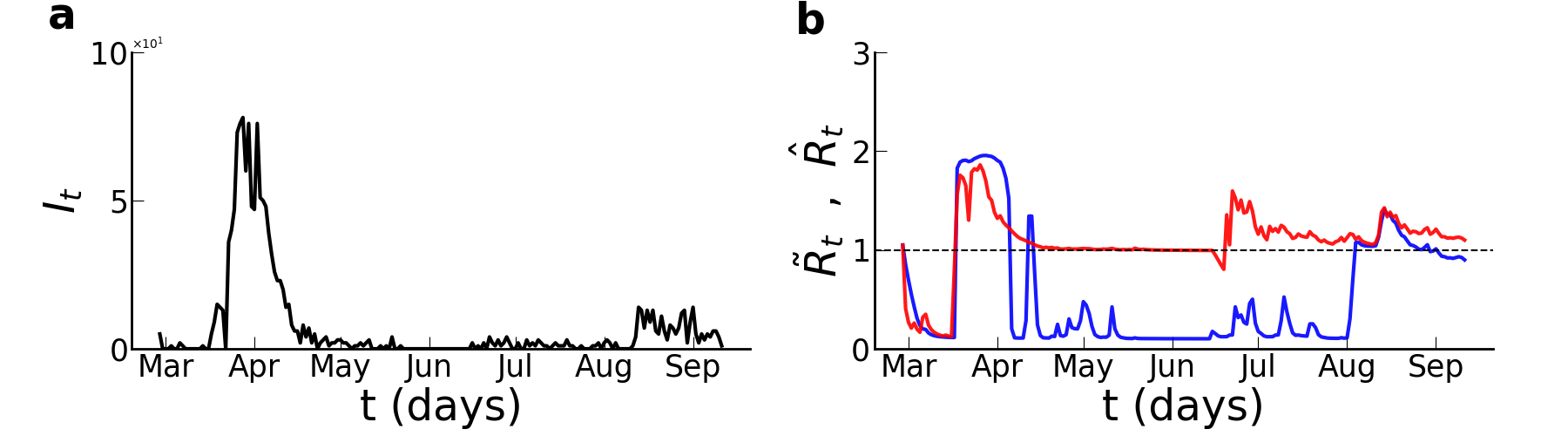}
  \caption{Temporal evolution of new infections of the COVID-19 in New Zealand (a) and their estimates by the EpiFilter method (red line) and our approach (blue line).}
  \label{realnz}
\end{figure*}

Governor Andrew Cuomo of New York declared a state of emergency on March 7th and progressively implemented stricter restrictions beginning March 16th. A stay-at-home order was issued on April 6th, and school closures were extended until April 29th. April 6th can be considered a pivotal moment, as $R_t$ below 1. When considering a time window surrounding this date, Epifiter's estimates initially rose before declining, with $\tilde{R}_t$ reaching $1$ on April 16th. Similarly, our method exhibited a trend of increase followed by decrease, with $\hat{R}_t$ dropping to $1$ on April 11th. Therefore, our approach shows greater accuracy in this context.

As a second example, we examine COVID-19 transmission patterns in New Zealand using incidence data from February 28th, 2020 to September 11th, 2020~\cite{wto}. New Zealand serves as an insightful case study because officials implemented swift lockdowns combined with extensive testing, successfully achieving and maintaining very low incidence levels that ultimately led to the local elimination of COVID-19~\cite{cousins2020}. In Fig.~\ref{realnz}, we present a comparison of the effective reproduction number estimates derived from the EpiFilter method ($\tilde{R}_t$) and our approach ($\hat{R}_t$). These estimates were calculated using Eqs.~\eqref{obeq} and~\eqref{state} for the EpiFilter method, and Eqs.~\eqref{ob3} and~\eqref{stateq} for our approach. Notably, there are substantial differences in the inference quality between $\tilde{R}_t$ and $\hat{R}_t$ that warrant attention.

While both methods capture the suppression of the initial wave in New Zealand in April, linked to the implementation of critical interventions, including the lockdown on April 1st~\cite{jefferies2020}, the generalized Epifliter more accurately estimate that the effective reproduction number ($\hat{R}_t$) dropped sharply below 1 starting April 6th. In contrast, the standard Epifliter estimate ($\tilde{R}_t$) only slowly decreased to 1 by April 11th. This suggests that our generalized Epifliter offers a more precise estimation of the effective reproduction number. In August, when new community transmission cases emerged in Auckland, our methodology correctly identify a rise in the effective reproduction number above 1 during this period, whereas the Epifliter fails to capture this change.

\section*{Conclussion}

The real-time estimation of the effective reproduction number $R_t$ is crucial for pandemic response and control. It not only assists decision-makers in timely adjusting intervention measures but also provides essential information to the public, guiding them to take appropriate protective behaviours to reduce the risk of being infected. In the real world, however, there are many factors make estimation challenging such as transmission heterogeneity~\cite{lloyd2005} and reporting delays~\cite{althouse2020}. Transmission heterogeneity involves variations in individuals' reproduction during the transmission process, while reporting delays involve differences in individuals' perception and reporting of events or information.

To capture the intricate dynamics of epidemic transmission, encompassing variations in transmission rates and reporting delays, we have introduced negative binomial and Weibull distributions to model these factors and derived the observation equation~\eqref{ob3}. For two hidden states $R_t$ and $r_t$, we assumed that they are autocorrelated and independent of each other, and obtained the state equation~\eqref{stateq}. Based on these two equations, the state-space model was formulated. We devised an exact filtering solution (Eq.~\eqref{filter2}) to estimate $R_t$.

To evaluate its effectiveness, we applied the aforementioned framework to both synthetic and empirical datasets. Through synthetic data examples, we effectively showcased the method's superiority in handling data characterized by varying transmission rates and reporting delays. By comparing the effective reproduction number estimates obtained with and without accounting for transmission heterogeneity or reporting delays, we substantiated the substantial improvement offered by our method over conventional approaches. When applied to real-world data, specifically the COVID-19 outbreak in New York State and New Zealand, our method accurately inferred $R_t$. Notably, our $R_t$ estimate closely aligns with the timing of intervention implementation. Thus, we contribute a fresh perspective to understanding the impact of individual differences in epidemic inference.

It is important to highlight that in our study, we have employed a fixed serial interval distribution. However, in a broader context, the serial interval distributions can vary across different stages of disease propagation. In addition, we have assumed conditional independence between observed and state values to construct the state-space model, adhering to the Markov assumption. This means that the current observed value $I_t$ relies solely on the current state value $R_t$. Yet, in reality, the observed values might depend on several preceding state values, indicating non-Markovian characteristics. Hence, for future investigations, it would be advantageous to consider these aspects and extend the framework further to achieve a more comprehensive grasp of the dynamic changes in the effective reproduction number.

\section*{Code availibility}

The code and data to implement the model are publicly available on GitHub at https://github.com/Huyssenwork/TVRN.

\section*{Author contributions}

X.-J.X. and L.-J.Z. conceived and developed the approach presented in the paper. S.-J.H. run the numerical
simulations. X.-J.X. wrote the paper. X.-J.X., S.-J.H., and L.-J.Z. analysed the results. All authors reviewed the manuscript.

\section*{Funding}

This work was supported by the Natural Science Foundation of China under Grant No. 12071281 and the Science and Technology Commission of Shanghai Municipality under Grant No. 22JC1401401.

\section*{Competing interests}

The authors declare no competing interests.



\end{document}